\documentclass[conference]{IEEEtran}

\usepackage{amsmath}                        
\usepackage{booktabs}                       
\usepackage{graphicx}
\usepackage{amssymb, subfigure, stfloats, amssymb, amsfonts, rotating, acronym}
\usepackage{multirow}
\usepackage{relsize}
\usepackage{array}
\usepackage{tabularx}
\usepackage{tikz}
\usetikzlibrary{patterns}
\usetikzlibrary{arrows,positioning,shapes.geometric,spy}
\usepackage{filecontents}
\usepackage{xcolor}
\usepackage{pgfplots}
\usepackage{cite}
\usepackage{adjustbox}
\usepackage{colortbl}
\usepackage{lipsum}
\DeclareMathOperator*{\sgn}{sgn}

\newcommand{\fixme}[2]{\ifx&#2&{\leavevmode\color{red}#1}\else{\leavevmode\color{red}FIXME\{}#1{\leavevmode\color{red}\}}\footnote{{\leavevmode\color{red}#2}}\PackageWarning{Fixme}{#1: #2}\fi}

\begin{document}

\title{On Error-Correction Performance and Implementation of Polar Code List Decoders for 5G}

\author{\IEEEauthorblockN{Furkan Ercan, Carlo Condo, Seyyed Ali Hashemi, Warren J. Gross}
\IEEEauthorblockA{Department of Electrical and Computer Engineering, McGill University, Montr\'eal, Qu\'ebec, Canada\\
Email: furkan.ercan@mail.mcgill.ca, carlo.condo@mail.mcgill.ca, seyyed.hashemi@mail.mcgill.ca, warren.gross@mcgill.ca}}

\maketitle

\begin{abstract}
Polar codes are a class of capacity achieving error correcting codes that has been recently selected for the next generation of wireless communication standards (5G). Polar code decoding algorithms have evolved in various directions, striking different balances between error-correction performance, speed and complexity. Successive-cancellation list (SCL) and its incarnations constitute a powerful, well-studied set of algorithms, in constant improvement. At the same time, different implementation approaches provide a wide range of area occupations and latency results. 5G puts a focus on improved error-correction performance, high throughput and low power consumption: a comprehensive study considering all these metrics is currently lacking in literature. In this work, we evaluate SCL-based decoding algorithms in terms of error-correction performance and compare them to low-density parity-check (LDPC) codes. Moreover, we consider various decoder implementations, for both polar and LDPC codes, and compare their area occupation and power and energy consumption when targeting short code lengths and rates. Our work shows that among SCL-based decoders, the partitioned SCL (PSCL) provides the lowest area occupation and power consumption, whereas fast simplified SCL (Fast-SSCL) yields the lowest energy consumption. Compared to LDPC decoder architectures, different SCL implementations occupy up to 17.1$\times$ less area, dissipate up to 7.35$\times$ less power, and up to 26$\times$ less energy.

\end{abstract}

\IEEEpeerreviewmaketitle

\section{Introduction}\label{introduction}
\IEEEPARstart{P}{olar} codes, introduced by Ar{\i}kan in \cite{arikan09}, are a class of error-correcting codes that can provably achieve channel capacity on a memoryless channel when the code length $N$ tends to infinity. They have been selected for the next generation of wireless communication standards \cite{3GPP-5G}. 

The 5G standardization process is putting a particular focus on improved error-correction performance, lower power consumption and higher throughput. For example, machine-to-machine communications in 5G target massive connectivity among a high number of devices, on a scale higher than the most bandwidth-demanding applications in 3G and 4G \cite{M2M-zheng14}, with a limited power budget. Therefore, reliable and efficient encoding and decoding methods need to be designed.

In \cite{arikan09}, the successive-cancellation (SC) decoding algorithm is proposed for polar codes: it can be represented as a binary tree search. While optimal with infinite code length, this approach suffers from long decoding latency and mediocre error-correction performance at moderate code lengths.
To improve the error-correction performance of SC, the SC List (SCL) decoding algorithm was proposed in \cite{TalList}, that relies on a list of $L$ codeword candidates. A cyclic-redundancy check (CRC) is also concatenated to the polar code, to help in the selection of the correct candidate at the end of the SCL decoding process. The improved error-correction performance of CRC-aided SCL comes at the cost of additional computational complexity and latency. A hardware implementation for SCL using logarithmic likelihood ratio (LLR) values was presented in \cite{list-LLR}. In order to reduce latency and increase throughput, simplified SCL (SSCL) \cite{SSCLconf16} and Fast-SSCL \cite{fastSSCL-jour} decoding algorithms were proposed, that rely on the identification of bit patterns to prune the SC decoding tree and reduce the number of required bit estimations, with minor or no error-correction performance degradation. Compared to the conventional SCL, SSCL and Fast-SSCL can reduce the number of time steps required to decode one codeword up to $88\%$ \cite{fastSSCL-jour}. To address the high implementation complexity of SCL decoders, a partitioned SCL (PSCL) decoder was proposed in \cite{PSCL-ICASSP}: it shows substantial area occupation reduction and negligible error-correction performance loss with respect to conventional SCL decoders.

SCL-based decoders are currently one of the best candidates to meet 5G error-correction performance requirements and throughput. While most recent decoder architectures for polar codes focus on improving throughput and area occupation, little work has been done in terms of power consumption \cite{lowpower_bp,cassagne_sw_polar}. A large part of machine-to-machine connected devices are mobile-end platforms that use batteries and small-scale energy harvesting electronics: ultra-low power/energy consumption for these devices is crucial \cite{5G-M2M-book}.

This work provides an extensive study on polar code SCL-based decoders in terms of frame error rate (FER) performance, area occupation, and power/energy consumption. We focus on short to medium code lengths, similar to those chosen for the eMBB control channel \cite{3gpp_shortCode}. For rates $\frac{1}{2}$ and $\frac{2}{3}$, SCL-based decoders are compared against low-density parity-check (LDPC) codes from the IEEE 802.16e (WiMAX) standard with variable maximum iteration number. Then, we address power consumption of polar code decoders based on SCL, SSCL, Fast-SSCL and PSCL, and compare them against LDPC codes.

The rest of this work is organized as follows. In Section~\ref{background}, polar codes are briefly introduced along with various SCL-based decoding algorithms. Hardware implementations of polar code decoders are discussed in Section~\ref{sec:hardware}. In Section~\ref{sec:compare}, the error-correction performance of polar codes is analyzed and compared to that of LDPC codes from communications standards. Section~\ref{sec:implresults} presents synthesis results for a wide variety of decoder architectures, and compares them to LDPC decoders in literature. Conclusions are drawn in Section~\ref{sec:conclusion}.

\section{Background}\label{background}

\subsection{Polar Codes}
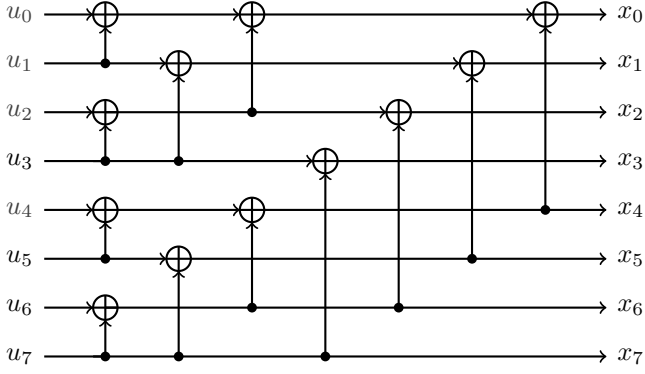
\begin{figure}
  \centering
  \begin{tikzpicture}[scale=.65, thick]
  \node [color=darkgray] at (.5,0) {$u_0$} ;
  \node [color=darkgray]at (.5,-1) {$u_1$};
  \node [color=darkgray]at (.5,-2) {$u_2$};
  \node at (.5,-3) {$u_3$};
  \node [color=darkgray]at (.5,-4) {$u_4$};
  \node at (.5,-5) {$u_5$};
  \node at (.5,-6) {$u_6$};
  \node at (.5,-7) {$u_7$};

  \foreach \x in {-6,-4,-2,0}
  {
    \draw [->] (1,\x) -- (2,\x);
    \draw (1,\x-1) -- (2.25,\x-1);

    \draw (2.25,\x) circle [radius=.25];
    \draw (2,\x) -- (2.5,\x);
    \draw (2.25,\x-.25) -- (2.25,\x+.25);

    \draw [->] (2.25,\x-1) -- (2.25,\x-.25);

    \fill (2.25,\x-1) circle [radius=.1];
  }

  \foreach \x in {-4,0}
  {
    \draw [->] (2.5,\x) -- (5,\x);
    \draw [->] (2.25,\x-1) -- (3.5,\x-1);

    \draw (5.25,\x) circle [radius=.25];
    \draw (5,\x) -- (5.5,\x);
    \draw (5.25,\x-.25) -- (5.25,\x+.25);

    \draw (3.75,\x-1) circle [radius=.25];
    \draw (3.5,\x-1) -- (4,\x-1);
    \draw (3.75,\x-1-.25) -- (3.75,\x-1+.25);

    \draw [->] (2.25,\x-2) -- (5.25,\x-2) -- (5.25,\x-.25);
    \fill (5.25,\x-2) circle [radius=.1];
    \draw [->] (2,\x-3) -- (3.75,\x-3) -- (3.75,\x-1-.25);
    \fill (3.75,\x-3) circle [radius=.1];
  }

  \draw [->] (5.5,0) -- (11,0);
  \draw [->] (4,-1) -- (9.5,-1);
  \draw [->] (5.25,-2) -- (8,-2);
  \draw [->] (3.75,-3) -- (6.5,-3);

  \foreach \x in {-1,0}
  {
    \draw [->] (5.5+1.5*\x,\x-4) -- (11.25+1.5*\x,\x-4) -- (11.25+1.5*\x,\x-.25);
    \draw [->] (5.25+1.5*\x,\x-6) -- (11.25+1.5*\x-3,\x-6) -- (11.25+1.5*\x-3,\x-2-.25);
  }

  \foreach \x in {-3,...,0}
  {
    \draw (11.25+1.5*\x,\x) circle [radius=.25];
    \draw (11+1.5*\x,\x) -- (11.5+1.5*\x,\x);
    \draw (11.25+1.5*\x,\x-.25) -- (11.25+1.5*\x,\x+.25);

    \fill (11.25+1.5*\x,\x-4) circle [radius=.1];

    \draw [->] (11.5+1.5*\x,\x) -- (12.5,\x);
    \draw [->] (11.25+1.5*\x,\x-4) -- (12.5,\x-4);
  }

  \node at (13,0) {$x_0$};
  \node at (13,-1) {$x_1$};
  \node at (13,-2) {$x_2$};
  \node at (13,-3) {$x_3$};
  \node at (13,-4) {$x_4$};
  \node at (13,-5) {$x_5$};
  \node at (13,-6) {$x_6$};
  \node at (13,-7) {$x_7$};

\end{tikzpicture}
  \caption{Polar code encoding for $PC(8,4)$. Gray indices indicate frozen bits while black indices represent information bits.}
  \label{fig:polarencode}
\end{figure}

Polar codes are able to achieve channel capacity through channel polarization, that splits $N$ channel utilizations into $K$ reliable ones, through which information bits are sent, and $N-K$ unreliable ones, used for frozen bits. A polar code, represented as $PC(N,K)$, is a linear block code of length $N = 2^n$ and rate $R = K/N$. Encoding of a polar code can be represented by a matrix multiplication:
\begin{equation}\label{eq:enc}
\boldsymbol{x_0^{N-1}} = \boldsymbol{u_0^{N-1}}G^{\otimes n}\text{,}
\end{equation}
where $\boldsymbol{u_0^{N-1}} = \{u_0,u_1,\ldots,u_{N-1}\}$ is the input vector, $\boldsymbol{x_0^{N-1}} = \{x_0,x_1,\ldots,x_{N-1}\}$ is the encoded vector, and the generator matrix $G^{\otimes n}$ is the $n$-th Kronecker product of the polar code matrix $G = \left[\begin{smallmatrix} 1&0\\ 1&1 \end{smallmatrix} \right]$. A polar code of length $N$ is composed of two concatenated polar codes of length $N/2$; Fig. \ref{fig:polarencode} depicts the encoding process for $PC(8,4)$.

In \cite{arikan09}, it was shown that as $N \rightarrow \infty$, encoded bits become either completely unreliable or completely reliable. For a polar code of rate $R=K/N$, $N-K$ most unreliable bits are fixed to a constant that is known by the decoder, usually to zero; remaining $K$ reliable locations are used to transmit the information bits. For the $PC(8,4)$ code in Fig. \ref{fig:polarencode}, bits $u_0$, $u_1$, $u_2$, and $u_4$ are located on the least reliable indices, thus are frozen and indicated with set $\Phi$ (gray indices in the figure), while bits $u_3$, $u_5$, $u_6$, and $u_7$ are located on the most reliable indices, which carry the information bits (black indices in the figure).

By its serial nature, SC decoding estimates a bit $\hat{u}_i$ according to the channel output $\boldsymbol{y_0^{N-1}} = \{y_0,y_1,\ldots,y_{N-1}\}$ and previously estimated bits $\boldsymbol{\hat{u}_0^{i-1}} = \{u_0,u_1,\ldots,u_{i-1}\}$. Let us represent the LLR value of $u_i$ as $\alpha_{u_i} = \ln \frac{\text{Pr}[\boldsymbol{y_0^{N-1}}, \boldsymbol{\hat{u}_0^{i-1}} | \hat{u}_i  = 0]}{\text{Pr}[\boldsymbol{y_0^{N-1}}, \boldsymbol{\hat{u}_0^{i-1}} | \hat{u}_i  = 1]}$. SC estimates each bit in accordance with
\begin{equation}\label{eqn:bitestimate-sc}
\hat{u}_{i}=\left\{
  \begin{array}{@{}ll@{}}
    0, & \text{when } \alpha_{u_i} \geq 0 \text{ or } i \in \Phi; \\
    1, & \text{otherwise.}
  \end{array}\right.
\end{equation}

\begin{figure}
  \centering
   \scalebox{0.75}{\begin{tikzpicture}[scale=.65, thick]

\draw [thin,gray,dashed] (0,-1) -- (7.5,-1);
\draw [thin,gray,dashed] (0,-3) -- (11.5,-3);
\draw [thin,gray,dashed] (0,-5) -- (13.5,-5);
\draw [thin,gray,dashed] (0,-7) -- (14.5,-7);

\node at (-.75,-1) {$S=3$};
\node at (-.75,-3) {$S=2$};
\node at (-.75,-5) {$S=1$};
\node at (-.75,-7) {$S=0$};

\draw (7.5,-1) -- (3.5,-3);
\draw (7.5,-1) -- (11.5,-3);

\draw (3.5,-3) -- (1.5,-5);
\draw (3.5,-3) -- (5.5,-5);
\draw (11.5,-3) -- (9.5,-5);
\draw (11.5,-3) -- (13.5,-5);

\draw (1.5,-5) -- (0.5,-7);
\draw (1.5,-5) -- (2.5,-7);

\draw (5.5,-5) -- (4.5,-7);
\draw (5.5,-5) -- (6.5,-7);

\draw (9.5,-5) -- (8.5,-7);
\draw (9.5,-5) -- (10.5,-7);

\draw (13.5,-5) -- (12.5,-7);
\draw (13.5,-5) -- (14.5,-7);


  \draw[black,fill=white] (.5,-7) circle [radius=.25];			
  \fill[color=gray] (1.5,-5) circle [radius=.25];	
  \draw[black,fill=white] (2.5,-7) circle [radius=.25];			
  \fill[color=gray] (3.5,-3) circle [radius=.25];	
  \draw[black,fill=white] (4.5,-7) circle [radius=.25];			
  \fill[color=gray] (5.5,-5) circle [radius=.25];	
  \fill[color=black] (6.5,-7) circle [radius=.25];	

  \fill[color=gray] (7.5,-1) circle [radius=.25];	

  \draw[black,fill=white] (8.5,-7) circle [radius=.25];		
  \fill[color=gray] (9.5,-5) circle [radius=.25];		
  \fill[color=black] (10.5,-7) circle [radius=.25];		
  \fill[color=gray] (11.5,-3) circle [radius=.25];		
  \fill[color=black] (12.5,-7) circle [radius=.25];		
  \fill[color=gray] (13.5,-5) circle [radius=.25];		
  \fill[color=black] (14.5,-7) circle [radius=.25];		

\node at (.5,-7.8) {$\hat{u}_0$};
\node at (2.5,-7.8) {$\hat{u}_1$};
\node at (4.5,-7.8) {$\hat{u}_2$};
\node at (6.5,-7.8) {$\hat{u}_3$};
\node at (8.5,-7.8) {$\hat{u}_4$};
\node at (10.5,-7.8) {$\hat{u}_5$};
\node at (12.5,-7.8) {$\hat{u}_6$};
\node at (14.5,-7.8) {$\hat{u}_7$};

\draw [->] (8,-1.125) -- (11,-2.625) node [above=.05cm,midway,rotate=-25] {$\boldsymbol{\alpha}$};
\draw [<-] (8,-1.375) -- (11,-2.875) node [below=-.05cm,midway,rotate=-25] {$\boldsymbol{\beta}$};

\draw [->] (11.25,-3) -- (9.75,-4.5) node [above=.03cm,midway,rotate=40] {$\boldsymbol{{\alpha}^l}$};
\draw [<-] (11.25,-3.5) -- (9.75,-5) node [below=-.05cm,midway,rotate=40] {$\boldsymbol{{\beta}^l}$};

\draw [->] (11.75,-3) -- (13.25,-4.5) node [above=.03cm,midway,rotate=-40] {$\boldsymbol{{\alpha}^r}$};
\draw [<-] (11.75,-3.5) -- (13.25,-5) node [below=-.05cm,midway,rotate=-40] {$\boldsymbol{{\beta}^r}$};

\end{tikzpicture}}
  \\
  \vspace{2pt}
  \caption{Succesive-cancellation decoding tree for a $PC(8,4)$ code.}
  \label{fig:scdecode}
\end{figure}
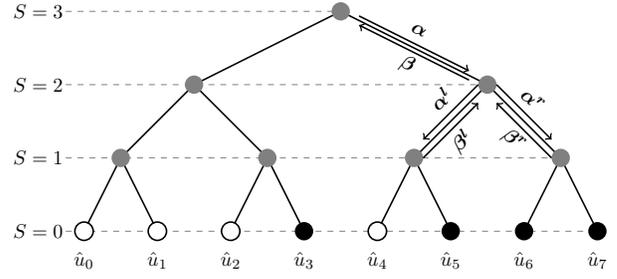

SC decoding traverses the polar code tree in Fig.~\ref{fig:scdecode} starting from the root node, and advances recursively from left to right. Each parent node at stage $S$ contains soft information (LLR values) $\boldsymbol{\alpha}=\{\alpha_0, \alpha_1, \ldots, \alpha_{2^S-1}\}$, and passes this soft information to its left and right children. Hard decision estimates $\boldsymbol{\beta}=\{\beta_0, \beta_1, \ldots, \beta_{2^S-1}\}$ are passed from child nodes to their parent nodes. From a parent node at stage $S$, the soft information passed to left child $\boldsymbol{\alpha^l} = \{\alpha^l_0, \alpha^l_1, \ldots, \alpha^l_{2^{S-1}-1}\}$ and right child $\boldsymbol{\alpha^r} = \{\alpha^r_0, \alpha^r_1, \ldots, \alpha^r_{2^{S-1}-1}\}$ can be approximated as
\begin{align}
{\alpha}^l_i &= \sgn(\alpha_{i})\sgn(\alpha_{i+2^{S-1}}) \min(\alpha_{i},\alpha_{i+2^{S-1}}) \text{,} \label{eqn:alphaleft}\\
{\alpha}^r_i &= \alpha_{i+2^{S-1}} + (1-2\beta^{l}_{i})\alpha_{i} \text{.} \label{eqn:alpharight}
\end{align}

The hard decision estimates $\boldsymbol{\beta}$ are calculated at each stage $S$ via the left and right messages received from child nodes, $\boldsymbol{\beta^{l}}=\{\beta^l_0, \beta^l_1, \ldots, \beta^l_{2^{S-1}-1}\}$ and $\boldsymbol{\beta^{r}}=\{\beta^r_0, \beta^r_1, \ldots, \beta^r_{2^{S-1}-1}\}$, as
\begin{equation}\label{eqn:beta}
  \beta_i=\left\{
  \begin{array}{@{}ll@{}}
    \beta^{l}_{i} \oplus \beta^{r}_{i}, & \text{if}~ i \leq 2^{S-1} \\
    \beta^{r}_{i}, & \text{otherwise.}
  \end{array}\right.
\end{equation}
where $\boldsymbol{\oplus}$ denotes bitwise XOR operation, and $0 \leq i < 2^S$. At the leaf nodes, $\beta$ values are hard decisions computed by (\ref{eqn:bitestimate-sc}). The computational complexity of SC decoding is $O(N\log_2 N)$.

\subsection{Successive-Cancellation List (SCL) Decoding}

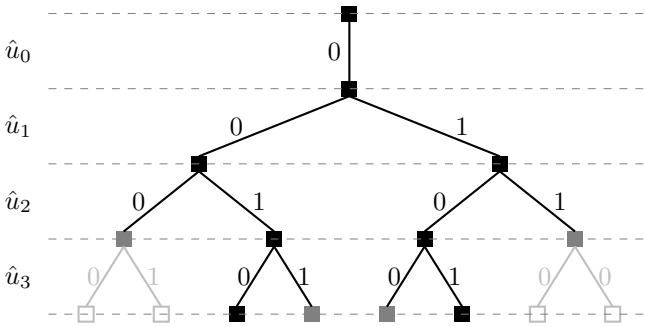
\begin{figure}
  \centering
  \hspace{-15pt}
   \scalebox{1}{\begin{tikzpicture}[scale=2, thick]
\newcommand\Square[1]{+(-#1,-#1) rectangle +(#1,#1)}

\fill (0,.5) \Square{.05};

\fill (0,0) \Square{.05};

\fill (-1,-.5) \Square{.05};
\fill (1,-.5) \Square{.05};

\fill [gray] (-1.5,-1) \Square{.05};
\fill (-.5,-1) \Square{.05};
\fill (.5,-1) \Square{.05};
\fill [gray] (1.5,-1) \Square{.05};

\draw [gray!50] (-1.75,-1.5) \Square{.05};
\draw [gray!50] (-1.25,-1.5) \Square{.05};
\fill (-.75,-1.5) \Square{.05};
\fill [gray] (-.25,-1.5) \Square{.05};
\fill [gray] (.25,-1.5) \Square{.05};
\fill (.75,-1.5) \Square{.05};
\draw [gray!50] (1.25,-1.5) \Square{.05};
\draw [gray!50] (1.75,-1.5) \Square{.05};

\draw (0,.45) -- (0,.05);

\draw (0,-.05) -- (-1,-.45);
\draw (0,-.05) -- (1,-.45);

\draw (-1,-.55) -- (-1.5,-.95);
\draw (-1,-.55) -- (-.5,-.95);
\draw (1,-.55) -- (.5,-.95);
\draw (1,-.55) -- (1.5,-.95);

\draw [gray!50] (-1.5,-1.05) -- (-1.75,-1.45);
\draw [gray!50] (-1.5,-1.05) -- (-1.25,-1.45);
\draw (-.5,-1.05) -- (-.75,-1.45);
\draw (-.5,-1.05) -- (-.25,-1.45);
\draw (.5,-1.05) -- (.25,-1.45);
\draw (.5,-1.05) -- (.75,-1.45);
\draw [gray!50] (1.5,-1.05) -- (1.25,-1.45);
\draw [gray!50] (1.5,-1.05) -- (1.75,-1.45);

\draw [very thin,gray,dashed] (-2,.5) -- (2,.5);
\draw [very thin,gray,dashed] (-2,0) -- (2,0);
\draw [very thin,gray,dashed] (-2,-.5) -- (2,-.5);
\draw [very thin,gray,dashed] (-2,-1) -- (2,-1);
\draw [very thin,gray,dashed] (-2,-1.5) -- (2,-1.5);

\node at (-2.2,.25) {$\hat{u}_0$};
\node at (-2.2,-.25) {$\hat{u}_1$};
\node at (-2.2,-.75) {$\hat{u}_2$};
\node at (-2.2,-1.25) {$\hat{u}_3$};

\node at (-.1,.25) {$0$};

\node at (-.75,-.25) {$0$};
\node at (.75,-.25) {$1$};

\node at (-1.4,-.75) {$0$};
\node at (-.6,-.75) {$1$};
\node at (.6,-.75) {$0$};
\node at (1.4,-.75) {$1$};

\node [gray!50] at (-1.7,-1.25) {$0$};
\node [gray!50] at (-1.3,-1.25) {$1$};
\node at (-.7,-1.25) {$0$};
\node at (-.3,-1.25) {$1$};
\node at (.3,-1.25) {$0$};
\node at (.7,-1.25) {$1$};
\node [gray!50] at (1.3,-1.25) {$0$};
\node [gray!50] at (1.7,-1.25) {$0$};

\end{tikzpicture}}
  \\
  \vspace{2pt}
  \caption{SCL decoding stages for list size $L = 2$.}
  \label{fig:listtree}
\end{figure}

In SCL decoding \cite{TalList}, when a bit is to be estimated, the decoding process splits into two paths; one path estimates the bit as a '$0$', and the other as a '$1$'. Therefore, at each bit estimation, the number of codeword paths double, until a list size $L$ is reached. In this context, SC can be considered as an SCL with list size $L=1$. Each path contains an information on the likelihood of the path being the correct codeword, which is defined as a path metric (PM). When the list size $L$ is doubled by estimating another bit in the sequence, the $L$ least likely paths are dropped based on their PM information, and the list is updated. Compared to SC, SCL decoding yields a better error-correction performance. Fig.~\ref{fig:listtree} depicts the parallel decoding process with list size $L=2$ for $PC(4,3)$: $\hat{u}_0$ is a frozen bit and as a result no path splitting occurs. Estimating $\hat{u}_1$ creates two paths with associated path reliability values. When $\hat{u}_2$ is estimated, out of four possible paths, two of them with least reliable PMs are discarded.

The PM is initialized as $0$, and at each bit estimation, PM is updated as 
\begin{equation}\label{eqn:pm}
	\text{PM}^i_{l}=\left\{
  \begin{array}{@{}ll@{}}
    \text{PM}^{i-1}_{l}, & \text{if}~ \hat{u}_i = \frac{1-\sgn(\alpha_{u_i})}{2} \text{,} \\
    \text{PM}^{i-1}_{l} + |\alpha_{u_i}|, & \text{otherwise,}
  \end{array}\right.
\end{equation}
where $l$ is the path index ($0 \leq l < L$), and $i$ is the estimated bit index.

In \cite{TalList}, it was observed that SCL decoding could pick a wrong codeword out of the final candidates if they are evaluated only by their PM, even when the correct codeword is present in the final list. Thus, a CRC is added as an outer decoding process to aid SCL decoding, which improves the error-correction performance significantly. On the other hand, SCL decoding suffers from long decoding latency and higher computational complexity of $O(LN\log_2 N)$. 




\subsection{Simplified SCL and Fast-SSCL Decoding}

The throughput of SC can be improved by an order of magnitude when applying the fast decoding techniques proposed in \cite{SSC2011} and \cite{sarkis14}. These techniques identify particular information and frozen bit patterns, reducing the decoding latency of SC with no error-correction performance degradation. Such special patterns are associated to nodes in the decoding tree: Rate-0 nodes (with no information bits), Rate-1 nodes (with no frozen bits), repetition (Rep) nodes (with a single information bit) and single parity-check (SPC) nodes (with a single frozen bit). In Fig.~\ref{fig:scdecode} the left and right child of the root node are examples of a Rep node and an SPC node, respectively. In \cite{sarkisfastlist}, it was shown that adaptation of these special nodes is applicable to SCL, yielding significant reduction in latency at the cost of error-correction performance loss.

The SSCL algorithm from \cite{SSCLconf16} proposes an efficient decoding technique by proving that Rate-0, Rate-1 and Rep nodes need not to be traversed to update the PM while guaranteeing error-correction performance preservation. This approach reduces the number of decoding steps for a node of length $N_v$ from $3N_v-2$, to $N_v$ for Rate-1 nodes, to $1$ for Rate-0 nodes, and to $2$ for Rep nodes \cite{hashemi_SSCL_TCASI}.

Fast-SSCL decoding \cite{fastSSCL-conf}, proposes an enhanced method to reduce the decoding latency further for Rate-1 nodes, down to $\min(L-1,N_v)$ time steps with zero error-correction performance degradation. It was shown that, when splitting the paths over the Rate-1 node, the path split that does not match the sign of the LLR will always be discarded after the $L-1$-th step.

\subsection{Partitioned SCL Decoding}

\begin{figure}
  \centering
\begin{tikzpicture}[scale=.65]
\usetikzlibrary{backgrounds}

\filldraw[fill=gray!40!white, draw=black] (+0.00,+0.00) circle [radius=.2];

\filldraw[fill=gray!40!white, draw=black] (-3.00,-1.00) circle [radius=.2];
\filldraw[fill=gray!40!white, draw=black] (+3.00,-1.00) circle [radius=.2];

\filldraw[fill=gray!40!white, draw=black] (-4.50,-2.20) circle [radius=.2];
\filldraw[fill=gray!40!white, draw=black] (-1.50,-2.20) circle [radius=.2];
\filldraw[fill=gray!40!white, draw=black] (+1.50,-2.20) circle [radius=.2];
\filldraw[fill=gray!40!white, draw=black] (+4.50,-2.20) circle [radius=.2];

\begin{scope}[on background layer]
\draw [-] (+0.00,+0.00) -- (-3.00,-1.00);
\draw [-] (+0.00,+0.00) -- (+3.00,-1.00);

\draw [-] (-3.00,-1.00) -- (-4.50,-2.20);
\draw [-] (-3.00,-1.00) -- (-1.50,-2.20);
\draw [-] (+3.00,-1.00) -- (+1.50,-2.20);
\draw [-] (+3.00,-1.00) -- (+4.50,-2.20);

\draw [-] (-4.50,-2.20) -- (-4.50,-2.70);
\draw [-] (-1.50,-2.20) -- (-1.50,-2.70);
\draw [-] (+1.50,-2.20) -- (+1.50,-2.70);
\draw [-] (+4.50,-2.20) -- (+4.50,-2.70);

\draw [dashed, color=gray] (-5.00,+0.00) -- (+0.00,-0.00);
\draw [dashed, color=gray] (-5.00,-1.00) -- (+3.00,-1.00);
\draw [dashed, color=gray] (-5.00,-2.20) -- (+4.50,-2.20);

\end{scope}

\node [color=black] at (-6.10,+0.00) {\scriptsize $S\!=\!\log_2 \! N$};
\node [color=black] at (-6.25,-1.00) {\scriptsize $S\!=\!\log_2 \! N\!-\!1$};
\node [color=black] at (-6.25,-2.20) {\scriptsize $S\!=\!\log_2 \! N\!-\!2$};

\draw [](-5.8,-2.70) rectangle ++(2.6,-1) node[pos=.5] {\scriptsize CRC-aided SCL};
\draw [](-2.8,-2.70) rectangle ++(2.6,-1) node[pos=.5] {\scriptsize CRC-aided SCL};
\draw [](+0.2,-2.70) rectangle ++(2.6,-1) node[pos=.5] {\scriptsize CRC-aided SCL};
\draw [](+3.2,-2.70) rectangle ++(2.6,-1) node[pos=.5] {\scriptsize CRC-aided SCL};

\end{tikzpicture}
  \\
  \vspace{2pt}
  \caption{Partitioned SCL decoding tree with $P=4$.}
  \label{fig:PSCLtree}
\end{figure}
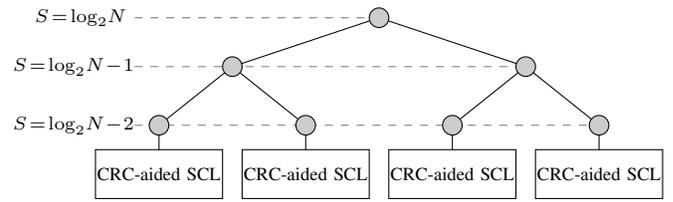

PSCL decoding divides the polar code into $P$ constituent sub-trees of length $N/P$, while every partition is decoded by the CRC-aided SCL algorithm \cite{PSCL-ICASSP}. Each partition has its own CRC, thus only one candidate is passed at the end of each partition to the next, using standard SC rules \cite{arikan09}. This approach helps reducing the memory requirements, since instead of storing $L$ copies of the complete tree, $L$ copies of a single partition are required. In addition, the same physical memory can be reused for different partitions. As a result, the memory requirements decrease exponentially with $P$. Fig.~\ref{fig:PSCLtree} depicts a generic PSCL decoder tree for a partition size of $P = 4$.

The reduced memory in PSCL comes at the cost of error-correction performance degradation compared to the conventional SCL decoding. As the number of partitions increases, the error-correction performance decays towards that of SC decoding. It was shown in \cite{PSCL-GLOBECOM} that a careful code construction and CRC selection can improve the error-correction performance of PSCL. 


\section{Hardware Architectures for SCL-Based Decoders}\label{sec:hardware}

\subsection{SCL Decoder}\label{sec:arch:base}

The architecture of the SCL decoder follows the one described in \cite{list-LLR}. It consists of five components: memory units, metric computation unit (MCU), metric sorting unit (MSU), address translation unit, and a controller. The MCU employs $L$ parallel SC decoders performing (\ref{eqn:alphaleft}), (\ref{eqn:alpharight}), and (\ref{eqn:beta}), one for each candidate codeword in the list. It also calculates the PM values whenever $L$ decision LLR values are calculated according to (\ref{eqn:pm}). It then takes one clock cycle to update and sort the PMs using MSU.
PMs are stored in a register-based memory architecture for each candidate, and are passed to a compute/swap unit at the end of each bit estimation. LLR and $\beta$ memory units have $L$ banks each, one for each parallel decoder unit. Considering there are $P_e$ processing elements available, each bank is itself divided into two parts, one handling the top stages of the decoding tree, where stage $S>\log_2(P_e)$, and one for the lower stages.

\subsection{SSCL \& Fast-SSCL Decoder}\label{sec:arch:sscl}

The architectures of SSCL and Fast-SSCL decoders are based on the SCL architecture described in Section~\ref{sec:arch:base}: they however expand MCU to perform Rate-0, Rate-1 and Rep node calculations. Size and position of special nodes in the decoder tree are computed offline and used by the decoder as inputs. For Rate-0, no path splitting occurs, and a single step is used to update the PM list. Rate-1 nodes are computed in two stages: first the portion of the information bits (all of them in SSCL) that are subject to path splitting is calculated. Then, in case of Fast-SSCL, the remaining bits are estimated in a single step, and their LLR values are used to update the PM according to (\ref{eqn:pm}). Computations for Rep nodes are similar to those of Rate-0 nodes: the frozen bits are treated as in Rate-0 nodes, while an additional step estimates the single information bit.

Both SSCL and Fast-SSCL architectures employ an $L$-parallel CRC computation unit that updates the CRC as soon as a bit is estimated by the SC decoders. They include different degrees of parallelism to accommodate the single-step estimation of multiple bits in Rate-0, Rate-1 and Rep nodes.

\subsection{PSCL Decoder}\label{sec:arch:pscl}

The PSCL decoder modifies the SCL decoder by reducing the size of the LLR and $\beta$ memories to fit the partition size. A single memory takes care of the top of the tree, where the SC rules are applied, and ad-hoc routing to processing elements is performed depending on the tree stage.

\section{Error Correction Performance Comparison}\label{sec:compare}
In this section, the error correction performances of SCL-based decoding algorithms described in Section~\ref{background} are evaluated and compared against each other, and against LDPC codes taken from the IEEE 802.16e standard where applicable. For polar codes, we consider code lengths of $256$ and $512$: these lengths are included in the 5G eMBB control channel \cite{3GPP-5G}. The LDPC code with $N=576$ has been instead selected from the WiMAX standard, being the only one of length comparable to that of polar codes. Our simulation environment considers additive white Gaussian noise (AWGN) channel and BPSK modulation.

The error-correction performance of SCL, SSCL and Fast-SSCL are identical. Therefore, they are referred with the notation SCL$L$-CRC$C$, where $L$ and $C$ denote the list size and the CRC length, respectively. PSCL decoders are referred as PSCL($P$,$L$)-CRC($c_0$,$c_1$,$\ldots$,$c_{P-1}$), where $P$ denotes the number of partitions and $c_p$ represents the CRC length of partition $p$. For LDPC codes, $T$ denotes the maximum number of iterations, and the normalized min-sum algorithm is used for decoding \cite{minsumLDPC}, together with layered scheduling \cite{Hocevar2004}.

The target code rates are $R \in \{\frac{1}{6}, \frac{1}{3}, \frac{1}{2}, \frac{2}{3}\}$ for polar codes, having been investigated in 5G discussions \cite{3gpp_shortCode}. Among these rates, WiMAX LDPC codes allow for $R \in \{\frac{1}{2}, \frac{2}{3}\}$.

A CRC of length $8$ is selected for polar codes. For PSCL, the CRC selection criteria from \cite{PSCL-GLOBECOM} was adopted. For a target $E_b/N_0$ value, a simulation sweeps the error-correction performance of PSCL with different CRC lengths. Only CRC polynomials of degrees which are multiples of four are considered, to reduce the algorithm complexity. Then, for each code length and rate, CRC lengths that provide the best error-correction performance are selected. 

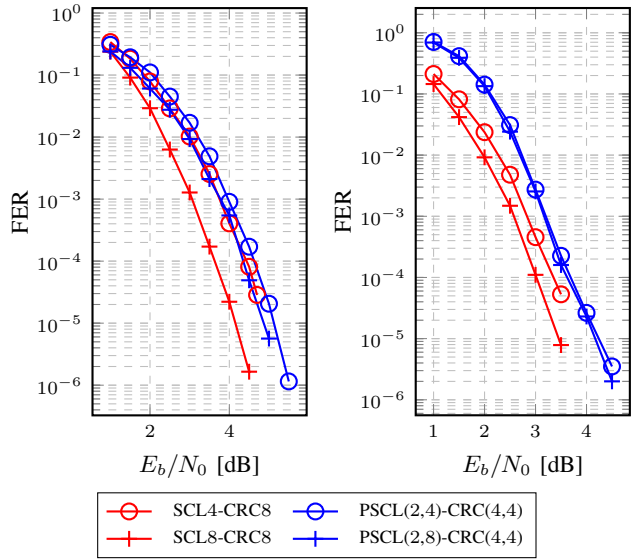
\begin{figure}
  \centering
\begin{tikzpicture}
  \pgfplotsset{
    label style = {font=\fontsize{9pt}{7.2}\selectfont},
    tick label style = {font=\fontsize{7pt}{7.2}\selectfont}
  }

\begin{axis}[
	scale = 1,
    ymode=log,
    xlabel={$E_b/N_0$ [\text{dB}]}, xlabel style={yshift=0.4em},
    ylabel={FER}, ylabel style={yshift=-0.75em},
    grid=both,
    ymajorgrids=true,
    xmajorgrids=true,
    grid style=dashed,
    width=0.5\columnwidth, height=7cm,
    thick,
    mark size=3,
    legend style={
      anchor={center},
      cells={anchor=west},
      column sep= 2mm,
      font=\fontsize{7pt}{7.2}\selectfont,
    },
    legend to name=SCL-n256-k42,
    legend columns=2,
]



\addplot[
    color=red,
    mark=o,
    mark size=3,
]
table {
1.0 3.40700e-01
1.5 1.83500e-01
2.0 7.89000e-02
2.5 2.90000e-02
3.0 1.02000e-02
3.5 2.53665e-03
4.0 4.06124e-04
4.5 8.10653e-05
4.7 2.85026e-05
};
\addlegendentry{SCL$4$-CRC$8$}

\addplot[
    color=blue,
    mark=o,
    mark size=3,
]
table {
1.0 3.08100e-01
1.5 1.93600e-01
2.0 1.10300e-01
2.5 4.50000e-02
3.0 1.70000e-02
3.5 4.89716e-03
4.0 9.04192e-04
4.5 1.70823e-04
5.0 2.05726e-05
5.5 1.14959e-06
};
\addlegendentry{PSCL($2$,$4$)-CRC($4$,$4$)}

\addplot[
    color=red,
    mark=+,
    mark size=3,
]
table {
1.0 2.38300e-01
1.5 9.10000e-02
2.0 2.92000e-02
2.5 6.30000e-03
3.0 1.27835e-03
3.5 1.71520e-04
4.0 2.21404e-05
4.5 1.64993e-06
};
\addlegendentry{SCL$8$-CRC$8$}

\addplot[
    color=blue,
    mark=+,
    mark size=3,
]
table {
1.0 2.40600e-01
1.5 1.30800e-01
2.0 6.07000e-02
2.5 2.76000e-02
3.0 9.30000e-03
3.5 2.10517e-03
4.0 5.45375e-04
4.5 4.90981e-05
5.0 5.64271e-06
};
\addlegendentry{PSCL($2$,$8$)-CRC($4$,$4$)}

\end{axis}
\end{tikzpicture}
\begin{tikzpicture}
  \pgfplotsset{
    label style = {font=\fontsize{9pt}{7.2}\selectfont},
    tick label style = {font=\fontsize{7pt}{7.2}\selectfont}
  }

\begin{axis}[
	scale = 1,
    ymode=log,
    xlabel={$E_b/N_0$ [\text{dB}]}, xlabel style={yshift=0.4em},
    ylabel={FER}, ylabel style={yshift=-0.75em},
    grid=both,
    ymajorgrids=true,
    xmajorgrids=true,
    grid style=dashed,
    width=0.5\columnwidth, height=7cm,
    thick,
    mark size=3,
    legend style={
      anchor={center},
      cells={anchor=west},
      column sep= 2mm,
      font=\fontsize{7pt}{7.2}\selectfont,
    },
    legend to name=SCL-n256-k42,
    legend columns=2,
]



\addplot[
    color=red,
    mark=o,
    thick,
    mark size=3,
]
table {
1.0 0.2117
1.5 0.0817
2.0 0.0237
2.5 0.00478194
3.0 0.000455419
3.5 5.30175e-05
};
\addlegendentry{SCL$4$-CRC$8$}

\addplot[
    color=blue,
    mark=o,
    thick,
    mark size=3,
]
table {
1.0 7.07500e-01
1.5 4.15000e-01
2.0 1.41900e-01
2.5 3.10000e-02
3.0 2.72985e-03
3.5 2.25758e-04
4.0 2.63573e-05
4.5 3.50913e-06
};
\addlegendentry{PSCL($2$,$4$)-CRC($4$,$4$)}

\addplot[
    color=red,
    mark=+,
    thick,
    mark size=3,
]
table {
1.0 0.1452
1.5 0.0418
2.0 0.0092
2.5 0.00148562
3.0 0.000110346
3.5 7.85123e-06
};
\addlegendentry{SCL$8$-CRC$8$}

\addplot[
    color=blue,
    mark=+,
    thick,
    mark size=3,
]
table {
1.0 6.92300e-01
1.5 3.89600e-01
2.0 1.34100e-01
2.5 2.42000e-02
3.0 2.54855e-03
3.5 1.59775e-04
4.0 2.30138e-05
4.5 2.00300e-06
};
\addlegendentry{PSCL($2$,$8$)-CRC($4$,$4$)}

\end{axis}
\end{tikzpicture}
  \\
  \vspace{2pt}
  \ref{SCL-n256-k42}
  \caption{FER of SCL and PSCL for polar codes with $N=256$ (left) and $N=512$ (right), and $R=1/6$.}
  \label{fig:SCL-R.16}
\end{figure}

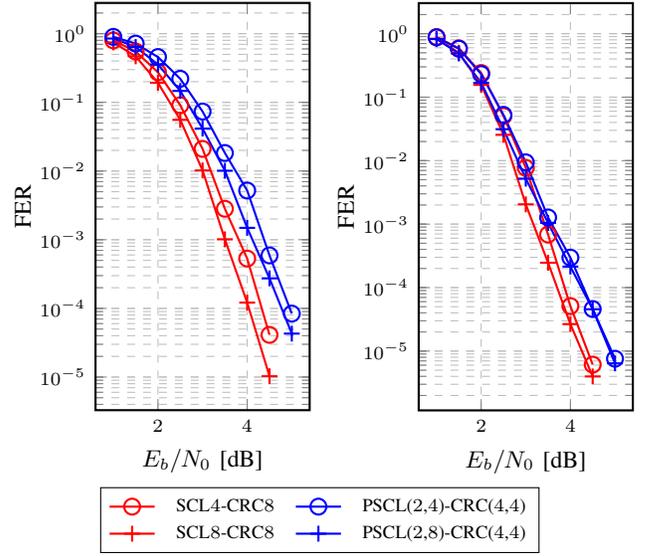
\begin{figure}
  \centering
\begin{tikzpicture}
  \pgfplotsset{
    label style = {font=\fontsize{9pt}{7.2}\selectfont},
    tick label style = {font=\fontsize{7pt}{7.2}\selectfont}
  }

\begin{axis}[
	scale = 1,
    ymode=log,
    xlabel={$E_b/N_0$ [\text{dB}]}, xlabel style={yshift=0.4em},
    ylabel={FER}, ylabel style={yshift=-0.75em},
    grid=both,
    ymajorgrids=true,
    xmajorgrids=true,
    grid style=dashed,
    width=0.5\columnwidth, height=7cm,
    thick,
    mark size=3,
    legend style={
      anchor={center},
      cells={anchor=west},
      column sep= 2mm,
      font=\fontsize{7pt}{7.2}\selectfont,
    },
    legend to name=SCL-n256-k128,
    legend columns=2,
]



\addplot[
    color=red,
    mark=o,
    thick,
    mark size=3,
]
table {
1.0 7.98400e-01
1.5 5.49900e-01
2.0 2.73900e-01
2.5 9.08000e-02
3.0 2.10000e-02
3.5 2.82215e-03
4.0 5.31107e-04
4.5 4.11592e-05
};
\addlegendentry{SCL$4$-CRC$8$}

\addplot[
    color=blue,
    mark=o,
    thick,
    mark size=3,
]
table {
1.0 8.98600e-01
1.5 7.17400e-01
2.0 4.56700e-01
2.5 2.20300e-01
3.0 7.35000e-02
3.5 1.83000e-02
4.0 5.20000e-03
4.5 5.90674e-04
5.0 8.38888e-05
};
\addlegendentry{PSCL($2$,$4$)-CRC($8$,$8$)}

\addplot[
    color=red,
    mark=+,
    thick,
    mark size=3,
]
table {
1.0 7.45300e-01
1.5 4.68100e-01
2.0 1.92800e-01
2.5 5.56000e-02
3.0 1.02000e-02
3.5 1.01317e-03
4.0 1.21277e-04
4.5 1.02739e-05
};
\addlegendentry{SCL$8$-CRC$8$}

\addplot[
    color=blue,
    mark=+,
    thick,
    mark size=3,
]
table {
1.0 8.49500e-01
1.5 6.40300e-01
2.0 3.56100e-01
2.5 1.46600e-01
3.0 4.15000e-02
3.5 1.01000e-02
4.0 1.48805e-03
4.5 2.73742e-04
5.0 4.29258e-05
};
\addlegendentry{PSCL($2$,$8$)-CRC($8$,$8$)}

\end{axis}
\end{tikzpicture}
\begin{tikzpicture}
  \pgfplotsset{
    label style = {font=\fontsize{9pt}{7.2}\selectfont},
    tick label style = {font=\fontsize{7pt}{7.2}\selectfont}
  }

\begin{axis}[
	scale = 1,
    ymode=log,
    xlabel={$E_b/N_0$ [\text{dB}]}, xlabel style={yshift=0.4em},
    ylabel={FER}, ylabel style={yshift=-0.75em},
    grid=both,
    ymajorgrids=true,
    xmajorgrids=true,
    grid style=dashed,
    width=0.5\columnwidth, height=7cm,
    thick,
    mark size=3,
    legend style={
      anchor={center},
      cells={anchor=west},
      column sep= 2mm,
      font=\fontsize{7pt}{7.2}\selectfont,
    },
    legend to name=SCL-n256-k128,
    legend columns=2,
]



\addplot[
    color=red,
    mark=o,
    thick,
    mark size=3,
]
table {
1.0 8.84500e-01
1.5 5.87100e-01
2.0 2.40300e-01
2.5 5.05000e-02
3.0 7.70000e-03
3.5 6.84922e-04
4.0 5.11525e-05
4.5 6.11373e-06
};
\addlegendentry{SCL$4$-CRC$8$}

\addplot[
    color=blue,
    mark=o,
    thick,
    mark size=3,
]
table {
1.0 8.74400e-01
1.5 5.79200e-01
2.0 2.30700e-01
2.5 5.26000e-02
3.0 9.40000e-03
3.5 1.27311e-03
4.0 2.96762e-04
4.5 4.59389e-05
5.0 7.57296e-06
};
\addlegendentry{PSCL($2$,$4$)-CRC($4$,$4$)}

\addplot[
    color=red,
    mark=+,
    thick,
    mark size=3,
]
table {
1.0 8.32600e-01
1.5 5.00900e-01
2.0 1.56200e-01
2.5 2.55000e-02
3.0 2.04876e-03
3.5 2.46354e-04
4.0 2.66180e-05
4.5 3.98532e-06
};
\addlegendentry{SCL$8$-CRC$8$}

\addplot[
    color=blue,
    mark=+,
    thick,
    mark size=3,
]
table {
1.0 8.22100e-01
1.5 4.87300e-01
2.0 1.68300e-01
2.5 3.13000e-02
3.0 5.20000e-03
3.5 1.04426e-03
4.0 2.14236e-04
4.5 4.56422e-05
5.0 6.40742e-06
};
\addlegendentry{PSCL($2$,$8$)-CRC($4$,$4$)}

\end{axis}
\end{tikzpicture}
  \\
  \vspace{2pt}
  \ref{SCL-n256-k128}
  \caption{FER of SCL and PSCL for polar codes with $N=256$ (left) and $N=512$ (right), and $R=2/3$.}
  \label{fig:SCL-R.66}
\end{figure}

Fig.~\ref{fig:SCL-R.16}, \ref{fig:SCL-R.33}, \ref{fig:SCL-R.5}, and \ref{fig:SCL-R.66} present the FER for SCL and PSCL algorithms with list sizes of $L \in \{4,8\}$, and code lengths of $N \in \{256, 512\}$ for code rates $1/6$, $1/3$, $1/2$, and $2/3$, respectively. A consistent improvement in FER can be seen when the list size is increased for all rates and lengths when SCL decoder is used. For a target FER of $10^{-4}$, this improvement reaches $1$ dB when a polar code of length $512$ with rate $1/3$ is used. Similar observations can be made in terms of PSCL, with a peak improvement of $0.25$ dB for $PC(512,256)$. In all cases, SCL decoders provide better error-correction performance than their PSCL counterparts.

\begin{figure}
  \centering
\begin{tikzpicture}
  \pgfplotsset{
    label style = {font=\fontsize{9pt}{7.2}\selectfont},
    tick label style = {font=\fontsize{7pt}{7.2}\selectfont}
  }

\begin{axis}[
	scale = 1,
    ymode=log,
    xlabel={$E_b/N_0$ [\text{dB}]}, xlabel style={yshift=0.4em},
    ylabel={FER}, ylabel style={yshift=-0.75em},
    grid=both,
    ymajorgrids=true,
    xmajorgrids=true,
    grid style=dashed,
    width=0.5\columnwidth, height=7cm,
    thick,
    mark size=3,
    legend style={
      anchor={center},
      cells={anchor=west},
      column sep= 2mm,
      font=\fontsize{7pt}{7.2}\selectfont,
    },
    legend to name=SCL-n256-k84,
    legend columns=2,
]



\addplot[
    color=red,
    mark=o,
    thick,
    mark size=3,
]
table {
1.0 2.15700e-01
1.5 8.05000e-02
2.0 2.03000e-02
2.5 5.00000e-03
3.0 5.46484e-04
3.5 5.73652e-05
};
\addlegendentry{SCL$4$-CRC$8$}

\addplot[
    color=blue,
    mark=o,
    thick,
    mark size=3,
]
table {
1.0 2.30800e-01
1.5 9.15000e-02
2.0 2.90000e-02
2.5 7.20000e-03
3.0 1.37031e-03
3.5 1.57271e-04
4.0 2.16105e-05
};
\addlegendentry{PSCL($2$,$4$)-CRC($4$,$4$)}

\addplot[
    color=red,
    mark=+,
    thick,
    mark size=3,
]
table {
1.0 1.29200e-01
1.5 4.44000e-02
2.0 6.60000e-03
2.5 1.10683e-03
3.0 1.47693e-04
3.5 2.02203e-05
4.0 2.83310e-06
};
\addlegendentry{SCL$8$-CRC$8$}

\addplot[
    color=blue,
    mark=+,
    thick,
    mark size=3,
]
table {
1.0 1.58200e-01
1.5 5.41000e-02
2.0 1.37000e-02 
2.5 3.77188e-03
3.0 6.45369e-04
3.5 8.35939e-05
4.0 2.31470e-05
};
\addlegendentry{PSCL($2$,$8$)-CRC($4$,$4$)}

\end{axis}
\end{tikzpicture}
\begin{tikzpicture}
  \pgfplotsset{
    label style = {font=\fontsize{9pt}{7.2}\selectfont},
    tick label style = {font=\fontsize{7pt}{7.2}\selectfont}
  }

\begin{axis}[
	scale = 1,
    ymode=log,
    xlabel={$E_b/N_0$ [\text{dB}]}, xlabel style={yshift=0.4em},
    ylabel={FER}, ylabel style={yshift=-0.75em},
    grid=both,
    ymajorgrids=true,
    xmajorgrids=true,
    grid style=dashed,
    width=0.5\columnwidth, height=7cm,
    thick,
    mark size=3,
    legend style={
      anchor={center},
      cells={anchor=west},
      column sep= 2mm,
      font=\fontsize{7pt}{7.2}\selectfont,
    },
    legend to name=SCL-n256-k42,
    legend columns=2,
]



\addplot[
    color=red,
    mark=o,
    thick,
    mark size=3,
]
table {
1.0 0.2114
1.5 0.0568
2.0 0.0101
2.5 0.000724575
3.0 3.03844e-05
};
\addlegendentry{SCL$4$-CRC$8$}

\addplot[
    color=blue,
    mark=o,
    thick,
    mark size=3,
]
table {
1.0 1.90200e-01
1.5 5.00000e-02
2.0 8.50000e-03
2.5 8.31283e-04
3.0 7.05725e-05
3.5 1.05626e-05
};
\addlegendentry{PSCL($2$,$4$)-CRC($4$,$4$)}

\addplot[
    color=red,
    mark=+,
    thick,
    mark size=3,
]
table {
1.0 1.50800e-01
1.5 2.57000e-02
2.0 2.67123e-03
2.5 1.46508e-04
3.0 4.62861e-06
};
\addlegendentry{SCL$8$-CRC$8$}

\addplot[
    color=blue,
    mark=+,
    thick,
    mark size=3,
]
table {
1.0 1.26900e-01
1.5 2.08000e-02
2.0 3.26371e-03
2.5 4.42752e-04
3.0 5.48457e-05
};
\addlegendentry{PSCL($2$,$8$)-CRC($4$,$4$)}

\end{axis}
\end{tikzpicture}
  \\
  \vspace{2pt}
  \ref{SCL-n256-k84}
  \caption{FER of SCL and PSCL for polar codes with $N=256$ (left) and $N=512$ (right), and $R=1/3$.}
  \label{fig:SCL-R.33}
\end{figure}

\begin{figure}
  \centering
\begin{tikzpicture}
  \pgfplotsset{
    label style = {font=\fontsize{9pt}{7.2}\selectfont},
    tick label style = {font=\fontsize{7pt}{7.2}\selectfont}
  }

\begin{axis}[
	scale = 1,
    ymode=log,
    xlabel={$E_b/N_0$ [\text{dB}]}, xlabel style={yshift=0.4em},
    ylabel={FER}, ylabel style={yshift=-0.75em},
    grid=both,
    ymajorgrids=true,
    xmajorgrids=true,
    grid style=dashed,
    width=0.5\columnwidth, height=7cm,
    thick,
    mark size=3,
    legend style={
      anchor={center},
      cells={anchor=west},
      column sep= 2mm,
      font=\fontsize{7pt}{7.2}\selectfont,
    },
    legend to name=SCL-n256-k128,
    legend columns=2,
]



\addplot[
    color=red,
    mark=o,
    thick,
    mark size=3,
]
table {
1.0 4.21300e-01
1.5 1.93800e-01
2.0 5.67000e-02
2.5 1.23000e-02
3.0 1.50630e-03
3.5 1.39381e-04
4.0 6.84327e-06
};
\addlegendentry{SCL$4$-CRC$8$}

\addplot[
    color=blue,
    mark=o,
    thick,
    mark size=3,
]
table {
1.0 3.84300e-01
1.5 1.63400e-01
2.0 6.08000e-02
2.5 1.34000e-02
3.0 3.55341e-03
3.5 5.41876e-04
4.0 1.28161e-04
4.5 3.21421e-05
};
\addlegendentry{PSCL($2$,$4$)-CRC($4$,$4$)}

\addplot[
    color=red,
    mark=+,
    thick,
    mark size=3,
]
table {
1.0 3.23600e-01
1.5 1.21800e-01
2.0 3.07000e-02
2.5 4.89956e-03
3.0 5.05582e-04
3.5 3.57439e-05
};
\addlegendentry{SCL$8$-CRC$8$}

\addplot[
    color=blue,
    mark=+,
    thick,
    mark size=3,
]
table {
1.0 3.07700e-01
1.5 1.23400e-01
2.0 3.63000e-02
2.5 7.90000e-03
3.0 1.79837e-03
3.5 4.10233e-04
4.0 1.23810e-04
4.5 3.24629e-05
};
\addlegendentry{PSCL($2$,$8$)-CRC($4$,$4$)}

\end{axis}
\end{tikzpicture}
\begin{tikzpicture}
  \pgfplotsset{
    label style = {font=\fontsize{9pt}{7.2}\selectfont},
    tick label style = {font=\fontsize{7pt}{7.2}\selectfont}
  }

\begin{axis}[
	scale = 1,
    ymode=log,
    xlabel={$E_b/N_0$ [\text{dB}]}, xlabel style={yshift=0.4em},
    ylabel={FER}, ylabel style={yshift=-0.75em},
    grid=both,
    ymajorgrids=true,
    xmajorgrids=true,
    grid style=dashed,
    width=0.5\columnwidth, height=7cm,
    thick,
    mark size=3,
    legend style={
      anchor={center},
      cells={anchor=west},
      column sep= 2mm,
      font=\fontsize{7pt}{7.2}\selectfont,
    },
    legend to name=SCL-n256-k128,
    legend columns=2,
]



\addplot[
    color=red,
    mark=o,
    thick,
    mark size=3,
]
table {
1.0 4.24800e-01
1.5 1.35000e-01
2.0 2.56000e-02
2.5 2.12866e-03
3.0 1.62686e-04
3.5 8.30701e-06
};
\addlegendentry{SCL$4$-CRC$8$}

\addplot[
    color=blue,
    mark=o,
    thick,
    mark size=3,
]
table {
1.0 5.22400e-01
1.5 1.97800e-01
2.0 4.33000e-02
2.5 6.70000e-03
3.0 4.30689e-04
3.5 2.58015e-05
};
\addlegendentry{PSCL($2$,$4$)-CRC($8$,$8$)}

\addplot[
    color=red,
    mark=+,
    thick,
    mark size=3,
]
table {
1.0 3.12400e-01
1.5 8.22000e-02
2.0 9.80000e-03
2.5 6.93558e-04
3.0 3.82383e-05
3.5 3.07042e-06
};
\addlegendentry{SCL$8$-CRC$8$}

\addplot[
    color=blue,
    mark=+,
    thick,
    mark size=3,
]
table {
1.0 4.11900e-01
1.5 1.23400e-01
2.0 2.00000e-02
2.5 2.09363e-03
3.0 8.87696e-05
3.5 3.96586e-06
};
\addlegendentry{PSCL($2$,$8$)-CRC($8$,$8$)}

\end{axis}
\end{tikzpicture}
  \\
  \vspace{2pt}
  \ref{SCL-n256-k128}
  \caption{FER of SCL and PSCL for polar codes with $N=256$ (left) and $N=512$ (right), and $R=1/2$.}
  \label{fig:SCL-R.5}
\end{figure}

\begin{figure}
  \centering
\begin{tikzpicture}[spy using outlines=
	{rectangle, magnification=2, connect spies}]

\begin{axis}[
	scale = 1,
    ymode=log,
    xlabel={$E_b/N_0$ [\text{dB}]}, xlabel style={yshift=0.4em},
    ylabel={FER}, ylabel style={yshift=-0.45em},
    grid=both,
    ymajorgrids=true,
    xmajorgrids=true,
    grid style=dashed,
    width=1\columnwidth, height=7cm,
    thick,
    mark size=3,
    legend style={
      anchor={center},
      cells={anchor=west},
      column sep= 2mm,
      font=\fontsize{7pt}{7.2}\selectfont,
    },
    legend to name=SCL-LDPC,
    legend columns=2,
]

\addplot[
    color=red,
    mark=o,
    thick,
    mark size=3,
]
table {
1.0 8.84500e-01
1.5 5.87100e-01
2.0 2.40300e-01
2.5 5.05000e-02
3.0 7.70000e-03
3.5 6.84922e-04
4.0 5.11525e-05
4.5 6.11373e-06
};
\addlegendentry{SCL$4$-CRC$8$}

\addplot[
    color=red,
    mark=+,
    thick,
    mark size=3,
]
table {
1.0 8.32600e-01
1.5 5.00900e-01
2.0 1.56200e-01
2.5 2.55000e-02
3.0 2.04876e-03
3.5 2.46354e-04
4.0 2.66180e-05
4.5 3.98532e-06
};
\addlegendentry{SCL$8$-CRC$8$}

\addplot[
    color=blue,
    mark=o,
    thick,
    mark size=3,
]
table {
1.0 8.74400e-01
1.5 5.79200e-01
2.0 2.30700e-01
2.5 5.26000e-02
3.0 9.40000e-03
3.5 1.27311e-03
4.0 2.96762e-04
4.5 4.59389e-05
5.0 7.57296e-06
};
\addlegendentry{PSCL($2$,$4$)-CRC($4$,$4$)}

\addplot[
    color=blue,
    mark=+,
    thick,
    mark size=3,
]
table {
1.0 8.22100e-01
1.5 4.87300e-01
2.0 1.68300e-01
2.5 3.13000e-02
3.0 5.20000e-03
3.5 1.04426e-03
4.0 2.14236e-04
4.5 4.56422e-05
5.0 6.40742e-06
};
\addlegendentry{PSCL($2$,$8$)-CRC($4$,$4$)}

\addplot[
    color=black,
    mark=o,
    thick,
    mark size=3,
]
table {
1.0 9.96000e-01
1.5 9.32000e-01
2.0 6.39000e-01
2.5 2.16000e-01
3.0 2.88000e-02
3.5 1.75000e-03
4.0 8.79000e-05
4.5 8.90000e-06
5.0 7.00000e-07
};
\addlegendentry{LDPC $T=5$}

\addplot[
    color=black,
    mark=+,
    thick,
    mark size=3,
]
table {
1.0 9.82000e-01
1.5 8.23000e-01
2.0 4.07000e-01
2.5 8.19000e-02
3.0 6.05000e-03
3.5 2.65000e-04
4.0 1.85000e-05
4.5 2.60000e-06
5.0 2.00000e-07
};
\addlegendentry{LDPC $T=10$}

\addplot[
    color=black,
    mark=square,
    thick,
    mark size=3,
]
table {
1.0 9.70000e-01
1.5 7.61000e-01
2.0 3.28000e-01
2.5 5.48000e-02
3.0 3.52000e-03
3.5 1.59000e-04
4.0 1.38000e-05
4.5 2.00000e-06
5.0 2.00000e-07
};
\addlegendentry{LDPC $T=20$}


\coordinate (spypoint) at (axis cs:3.5,2e-4);
\coordinate (magnifyglass) at (axis cs:2.0,1e-5);

\end{axis}
\spy [blue, height=2.6cm, width=2cm] on (spypoint)
   in node[fill=white] at (magnifyglass);
\end{tikzpicture}
  \\
  \vspace{2pt}
  \ref{SCL-LDPC}
  \caption{FER for polar code with $N=512$, $R=2/3$ with SCL decoding compared against LDPC $N=576$, $R=2/3$ with various T.}
  \label{fig:SCL-LDPC-066}
\end{figure}
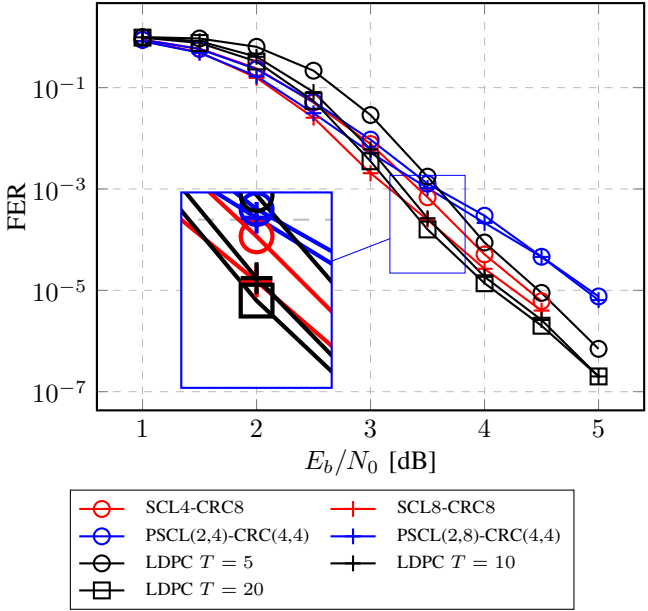

Fig.~\ref{fig:SCL-LDPC-05} and \ref{fig:SCL-LDPC-066} present the FER of polar codes with $N=512$ against LDPC WiMAX codes with $N=576$, for $R\in \{1/2,~2/3\}$. The maximum number of iterations considered for LDPC decoding is $T \in \{5, 10, 20\}$. For $R=1/2$ codes, SCL algorithm with $L=2$ outperforms LDPC with $T=20$, while at FER$=10^{-4}$, PSCL($2$,$2$)-CRC($8$,$8$) has the same FER. For $R=2/3$ codes in Fig.~\ref{fig:SCL-LDPC-066}, LDPC with $T=10$ matched the error-correction performance of SCL$8$-CRC$8$. Based on these results, in the following section we compare decoder architectures that target codes with matching FER. Thus, LDPC decoders are compared to PSCL for $R=2/3$, while SCL, SSCL and Fast-SSCL are used in case of $R=1/2$.

\begin{figure}
  \centering
\begin{tikzpicture}[spy using outlines=
	{rectangle, magnification=2, connect spies}]

\begin{axis}[
	scale = 1,
    ymode=log,
    xlabel={$E_b/N_0$ [\text{dB}]}, xlabel style={yshift=0.4em},
    ylabel={FER}, ylabel style={yshift=-0.45em},
    grid=both,
    ymajorgrids=true,
    xmajorgrids=true,
    grid style=dashed,
    width=1\columnwidth, height=7cm,
    thick,
    mark size=3,
    legend style={
      anchor={center},
      cells={anchor=west},
      column sep= 2mm,
      font=\fontsize{7pt}{7.2}\selectfont,
    },
    legend to name=SCL-LDPC-05,
    legend columns=2,
]

\addplot[
    color=red,
    mark=square,
    thick,
    mark size=3,
]
table {
1.0 5.77700e-01
1.5 2.51600e-01
2.0 7.09000e-02
2.5 1.05000e-02
3.0 1.31385e-03
3.5 8.89648e-05
};
\addlegendentry{SCL$2$-CRC$8$}

\addplot[
    color=red,
    mark=o,
    thick,
    mark size=3,
]
table {
1.0 4.24800e-01
1.5 1.35000e-01
2.0 2.56000e-02
2.5 2.12866e-03
3.0 1.62686e-04
3.5 8.30701e-06
};
\addlegendentry{SCL$4$-CRC$8$}


\addplot[
    color=blue,
    mark=square,
    thick,
    mark size=3,
]
table {
1.0 6.72300e-01
1.5 3.45500e-01
2.0 1.09600e-01
2.5 1.84000e-02
3.0 2.90867e-03
3.5 2.40036e-04
4.0 2.97689e-05
4.5 1.20031e-06
};
\addlegendentry{PSCL($2$,$2$)-CRC($8$,$8$)}

\addplot[
    color=blue,
    mark=o,
    thick,
    mark size=3,
]
table {
1.0 5.22400e-01
1.5 1.97800e-01
2.0 4.33000e-02
2.5 6.70000e-03
3.0 4.30689e-04 
3.5 2.58015e-05
};
\addlegendentry{PSCL($2$,$4$)-CRC($8$,$8$)}


\addplot[
    color=black,
    mark=o,
    thick,
    mark size=3,
]
table {
1.0 9.43000e-01
1.5 7.15000e-01
2.0 3.29000e-01
2.5 8.04000e-02
3.0 1.26000e-02
3.5 1.96000e-03
4.0 4.04000e-04
4.5 9.25000e-05
5.0 2.15000e-05
};
\addlegendentry{LDPC $T=5$}

\addplot[
    color=black,
    mark=+,
    thick,
    mark size=3,
]
table {
1.0 7.79000e-01
1.5 3.87000e-01
2.0 9.15000e-02
2.5 1.13000e-02
3.0 1.48000e-03
3.5 3.07000e-04
4.0 7.31000e-05
4.5 1.82000e-05
5.0 5.40000e-06
};
\addlegendentry{LDPC $T=10$}

\addplot[
    color=black,
    mark=square,
    thick,
    mark size=3,
]
table {
1.0 6.59000e-01
1.5 2.61000e-01
2.0 4.77000e-02
2.5 5.54000e-03
3.0 9.31000e-04
3.5 2.34000e-04
4.0 5.84000e-05
4.5 1.57000e-05
5.0 5.10000e-06
};
\addlegendentry{LDPC $T=20$}


\coordinate (spypoint) at (axis cs:3.5,2e-4);
\coordinate (magnifyglass) at (axis cs:1.5,5e-5);

\end{axis}
\spy [blue, height=2.6cm, width=2.2cm] on (spypoint)
   in node[fill=white] at (magnifyglass);
\end{tikzpicture}
  \\
  \vspace{2pt}
  \ref{SCL-LDPC-05}
  \caption{FER for polar code with $N=512$, $R=1/2$ with SCL decoding compared against LDPC $N=576$, $R=1/2$ with various T.}
  \label{fig:SCL-LDPC-05}
\end{figure}
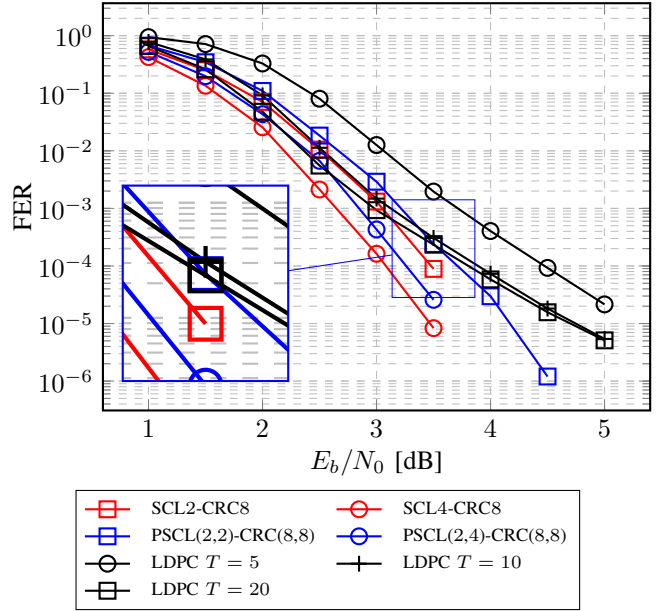

\section{ASIC Implementation Results}\label{sec:implresults}

In this section, synthesis results for SCL, SSCL, Fast-SSCL and PSCL for $N \in \{256,512\}$, $R \in \{{\frac{1}{2}, \frac{2}{3}}\}$, and $L \in \{2,4,8\}$ are presented. For each architecture, the number of parallel processing elements is $P_e = 32$. Based on simulations, PM quantization is selected as $8$ bits. For channel LLR and internal LLR values, quantization is $4$ and $6$ bits respectively, two of which are assigned to the fractional part. All memories have been synthesized as registers.

\begin{table}[t]
\centering
\caption{Synthesis area and energy consumption results with 65~nm TSMC CMOS technology for SCL, SSCL, Fast-SSCL and PSCL decoding of polar codes, $P_e=32$, and $f=800$~MHz.}
\label{allImpl}
\setlength{\extrarowheight}{1.5pt}
\begin{tabular}{c|c|c|c|c|ccc}
\toprule
\hline
\multirow{2}{*}{Algorithm} & \multirow{2}{*}{$N$} & \multirow{2}{*}{$R$} & \multirow{2}{*}{$L$} & CRC&  Area     & Power & Energy \\
	  &     &     &     &  [bits]   & [mm$^2$] & [mW] & [nJ]\\
\hline

\multirow{12}{*}{SCL} & \multirow{6}{*}{$256$}  & \multirow{3}{*}{$\frac{1}{2}$} &	$2$	& \multirow{12}{*}{$8$}  & $0.116$  & $35.99$ & $30.68$\\
			&		      &			     &	$4$	&  & $0.233$  & $83.85$ & $71.48$\\
			&		      &			     &	$8$	&  & $0.554$  & $192.52$ & $164.12$\\
\cline{3-4}
\cline{6-8}
			&		      & \multirow{3}{*}{$\frac{2}{3}$} &	$2$	&   & $0.116$  & $35.99$ & $32.62$\\
			&		      &			     &	$4$	&   & $0.233$  & $83.85$ & $75.99$\\			
			&		      &			     &	$8$	&   & $0.554$  &$192.52$ & $174.47$\\
\cline{2-4}
\cline{6-8}
		        & \multirow{6}{*}{$512$}& \multirow{3}{*}{$\frac{1}{2}$} &	$2$	&   & $0.215$  & $75.27$ & $128.54$\\
			&		      &			     &	$4$	&   & $0.432$  & $150.03$ & $256.21$\\			
			&		      &			     &	$8$	&   & $1.006$  & $345.39$ & $589.82$\\		        
\cline{3-4}
\cline{6-8}
			&		      & \multirow{3}{*}{$\frac{2}{3}$} &	$2$	&   & $0.215$  & $75.27$ & $136.65$\\		      
			&		      &			     &	$4$	&   & $0.432$  &$150.03$  & $272.38$\\			
			&		      &			     &	$8$	&   & $1.006$  & $345.39$ &$627.06$\\
\hline			
\multirow{12}{*}{SSCL} & \multirow{6}{*}{$256$} & \multirow{3}{*}{$\frac{1}{2}$} &	$2$	& \multirow{12}{*}{$8$} & $0.206$ & $56.77$& $23.99$ \\
			&		      &			     &	$4$	& & $0.395$    & $119.91$&$50.66$ \\			
			&		      &			     &	$8$	& & $0.826$    & $277.17$& $117.10$\\
\cline{3-4}
\cline{6-8}
			&		      & \multirow{3}{*}{$\frac{2}{3}$} &	$2$	& & $0.206$    &$56.77$ & $29.45$\\
			&		      &			     &	$4$	& & $0.395$    &$119.91$  &$62.20$\\
			&		      &			     &	$8$	& & $0.826$    &$277.17$  &$143.78$\\
\cline{2-4}
\cline{6-8}
			& \multirow{6}{*}{$512$}& \multirow{3}{*}{$\frac{1}{2}$} &	$2$	& & $0.343$     & $98.91$ &$83.70$\\
			&		      &			     &	$4$	& & $0.628$    & $192.97$ &$163.30$\\			
			&		      &			     &	$8$	& & $1.314$    &$421.47$  &$356.67$\\
\cline{3-4}
\cline{6-8}
			&		      & \multirow{3}{*}{$\frac{2}{3}$} &	$2$	& & $0.343$    & $98.91$ &$102.25$\\
			&		      &			     &	$4$	& & $0.628$    & $192.97$ &$199.48$\\			
			&		      &			     &	$8$	& & $1.314$    & $421.47$ &$435.69$\\			
\hline
\multirow{12}{*}{Fast-SSCL} & \multirow{6}{*}{$256$} & \multirow{3}{*}{$\frac{1}{2}$} &$2$	& \multirow{12}{*}{$8$} & $0.247$  & $65.85$& $14.90$ \\
			&		      &			     &	$4$	& & $0.490$    &$142.33$ & $41.28$\\			
			&		      &			     &	$8$	& & $1.049$    &$323.46$ & $109.17$\\
\cline{3-4}
\cline{6-8}
			&		      & \multirow{3}{*}{$\frac{2}{3}$} &	$2$	& &  $0.247$   &$65.85$ & $14.40$\\
			&		      &			     &	$4$	& &  $0.490$   &$142.33$ &$42.88$ \\			
			&		      &			     &	$8$	& &  $1.049$   &$323.46$ &$125.75$ \\
\cline{2-4}
\cline{6-8}
			& \multirow{6}{*}{$512$} & \multirow{3}{*}{$\frac{1}{2}$} &	$2$	& &  $0.422$   & $119.68$& $51.01$ \\
			&		      &			     &	$4$	& &  $0.795$   & $221.92$& $124.00$\\			
			&		      &			     &	$8$	& & $1.685$    & $493.43$& $341.08$\\
\cline{3-4}
\cline{6-8}
			&		      & \multirow{3}{*}{$\frac{2}{3}$} &	$2$	& &  $0.422$   & $119.68$& $45.18$ \\			
			&		      &			    &	$4$	 & & $0.795$   & $221.92$& $112.35$\\		
			&		      &			     &	$8$	& &  $1.685$   & $493.43$& $315.18$\\			
\hline			
\multirow{12}{*}{\begin{tabular}{@{}c@{}}PSCL \\ $P=2$\end{tabular}} & \multirow{6}{*}{$256$} & \multirow{3}{*}{$\frac{1}{2}$} &	$2$ &\multirow{12}{*}{($4$,$4$)} &  $0.091$   &$28.62$ & $24.40$\\
			&		      &			     &	$4$	& &  $0.164$   & $50.12$ & $42.73$\\			
			&		      &			     &	$8$	& &  $0.389$  &$123.55$ &$105.33$ \\
\cline{3-4}
\cline{6-8}
			&		      & \multirow{3}{*}{$\frac{2}{3}$} &	$2$	& &   $0.091$   & $28.62$& $25.94$\\
			&		      &			    &	$4$	 & &  $0.164$  & $50.12$& $45.42$\\		
			&		      &			    &	$8$	 & & $0.389$    & $123.55$& $111.97$ \\
\cline{2-4}
\cline{6-8}
			& \multirow{6}{*}{$512$} & \multirow{3}{*}{$\frac{1}{2}$}&	$2$	 & & $0.191$    & $63.19$ & $107.91$\\
			&		      &			     &	$4$	& &  $0.337$   & $119.53$ &$204.13$\\			
			&		      &			    &	$8$	 & & $0.694$    &$205.68$  &$351.25$\\
\cline{3-4}
\cline{6-8}
			&		      & \multirow{3}{*}{$\frac{2}{3}$}&	$2$	 &  & $0.191$   & $63.19$ &$114.72$\\		
			&		      &			    &	$4$	 & & $0.337$    & $119.53$ &$217.01$\\		
			&		      &			    &	$8$	 & & $0.694$    & $205.68$ &$373.41$\\		
\hline
\bottomrule
\end{tabular}
\end{table}

The architectures are synthesized with TSMC 65~nm CMOS technology, targeting a frequency of $f = 800$ MHz. Table~\ref{allImpl} compares the total area, power and energy consumption per codeword for all four SCL-based decoder implementations under the aforementioned design parameters.

The SCL decoder yields lower area occupation and power consumption compared to SSCL and Fast-SSCL, With all the considered code lengths, rates and list sizes. This is due to the fact that the special node computations in SSCL and Fast-SSCL add substantial logic complexity. Additional complexity is also caused by the parallel CRC units necessary to update Rate-0 and Rep nodes in SSCL, and also Rate-1 nodes in Fast-SSCL. In terms of energy consumption, Fast-SSCL provides the best results compared to its predecessors: although the power consumption is higher, the number of time steps needed to decode a codeword is reduced dramatically, yielding the lowest energy per frame.

In SCL-based implementations, memory is a major contribution in both area occupation and power consumption, that decreases exponentially with the partitioning factor of PSCL. Thus PSCL, with its reduced memory requirements, has the smallest area occupation and power consumption. In this context, with a minor degradation in performance, PSCL provides the best results for area- and power-efficient implementations.

For all considered rates in Table~\ref{allImpl} when $N = 256$, energy consumption per codeword of PSCL follows a close trend to SSCL when $L=2$. SCL, with its long decoding process, has the worst energy consumption, while Fast-SSCL has the lowest one. As $L$ increases, PSCL energy dissipation becomes comparable to that of Fast-SSCL. For $N = 512$, energy consumption for PSCL sits between that of SCL and SSCL for $L \in \{2,4\}$. For $L = 8$, the energy consumption of PSCL is lower than that of SSCL and higher than that of Fast-SSCL. This is due to the nonlinear increment in power consumption that both SSCL and Fast-SSCL experience with increasing $L$.

Table~\ref{tab:PC-LDPC} compares power, energy, and area of the considered polar code decoders against architectures for LDPC 802.16e codes taken from \cite{LDPCimpl1}, \cite{LDPCimpl2}, and \cite{LDPCimpl3}. Polar code decoders are selected based on the observations from Fig.~\ref{fig:SCL-LDPC-05}-\ref{fig:SCL-LDPC-066}. Note that the LDPC decoder architectures from \cite{LDPCimpl1} and \cite{LDPCimpl3} support both considered rates $R= \{ \frac{1}{2}, \frac{2}{3} \}$. Energy consumption for the LDPC architectures in Table~\ref{tab:PC-LDPC} are calculated with the number of iterations $T$ required to match the FER of polar codes from Section~\ref{sec:compare}.

\begin{table*}[t]
\centering
\caption{Comparative area, power, and energy consumption results for SCL, SSCL, Fast-SSCL and PSCL architectures for polar codes with $N=512$ against LDPC 802.16{e} architectures with $N=576$.}
\label{tab:PC-LDPC}
\begin{tabular}{l|c c c | c c c | c c c}
\toprule
\hline
 & SCL\textsuperscript{a} & Fast-SSCL\textsuperscript{a} & PSCL\textsuperscript{a,b} & SCL\textsuperscript{c} & SSCL\textsuperscript{c} & Fast-SSCL\textsuperscript{c} & LDPC\cite{LDPCimpl1} & LDPC\cite{LDPCimpl2} & LDPC\cite{LDPCimpl3} \\
 \hline
 Tech. (nm) & 65 & 65 & 65 & 65 & 65 & 65 & 90 & 180 & 90\\
 Rate & $1/2$ & $1/2$ & $1/2$ & $2/3$ & $2/3$ & $2/3$ & Any & $1/2$ & Any \\
 Area (mm$^2$) & $0.215$  & $0.422$  & $0.191$  & $1.006$  & $1.314$ & $1.685$ & $6.22$ & N/A & $6.25$\\
 Area\textsuperscript{d} (mm$^2$) & $0.215$  & $0.422$  & $0.191$  & $1.006$  & $1.314$ & $1.685$ & $3.24$ & N/A & $3.26$\\
 Power (mW)    & $75.27$  & $119.68$ & $63.19$  & $345.39$ & $421.47$ & $493.43$ & $528$  & $553$ & $264$\\
 Energy (nJ)   & $128.54$ & $51.01$  & $107.91$ & $589.82$ & $356.67$ & $315.18$ & $1368$ & $232.9$ & $690.6$\\
\hline
\bottomrule
 \multicolumn{2}{l}{\textsuperscript{a} $L=2$, $C=8$.} &
 \multicolumn{2}{l}{\textsuperscript{b} $P=2$, $(c_0,c_1)=(8,8)$.} &
 \multicolumn{2}{l}{\textsuperscript{c} $L=8$, $C=8$.} &
 \multicolumn{3}{l}{\textsuperscript{d} Scaled to 65 nm.} \\

\end{tabular}
\end{table*}

In Table~\ref{tab:PC-LDPC}, the area occupation for the LDPC decoders is scaled to 65~nm technology for a fair comparison. For rate $R=\frac{1}{2}$, the total area of polar code decoders ranges between $7.7\times$ (Fast-SSCL vs. \cite{LDPCimpl1}) to $17.1\times$ (PSCL vs. \cite{LDPCimpl3}) less than that of LDPC WiMAX implementations. For rate $R=\frac{2}{3}$, the advantage of polar decoders over LDPC decoders is lower, with a minimum of $2.46\times$ less area occupation.

Comparing power and energy consumption of architectures implemented with different technology nodes is not desirable, power scaling leads to wildly inaccurate figures. However, with the current scheme, SCL decoders consume up to $8.75\times$ less power, and up to $26.8\times$ less energy per frame in case of $R=\frac{1}{2}$. For $R=\frac{2}{3}$, polar codes yield $2.6\times$ less power consumption and $3.9\times$ less energy dissipation per frame.

According to these results, SCL-based polar code decoder implementations offer good solutions for 5G applications that require low area, power or energy consumption. For communication devices that require low power and energy, SCL, Fast-SSCL, and PSCL offer better figures than LDPC codes at comparable FER, code lengths and rates. Considering area occupation along with power consumption, PSCL provides a very favorable solution with negligible loss in error-correction performance.

\section{Conclusion}\label{sec:conclusion}
In this work, we evaluate SCL-based polar code decoder implementations in terms of error-correction performance, area occupation, power consumption, and energy consumption, for a code set case study. SCL, SSCL and Fast-SSCL have the same error-correction performance, while PSCL suffers minor FER loss. We show that the considered polar code decoders have comparable error-correction performance against WiMAX LDPC codes. We also show and compare the area, power and energy consumption for all four decoder implementations, and discuss their trade-offs. Comparing selected SCL-based decoder implementations against WiMAX LDPC architectures show that polar code decoders have reduced area, power and energy consumption, which makes them more suitable for potential 5G communications.



\end{document}